# Datenkompetenz im Physikstudium – ein Erfahrungsbericht

**Michael Krieger, Heiko B. Weber**
Lehrstuhl für Angewandte Physik, Department Physik, Friedrich-Alexander-Universität Erlangen-Nürnberg

**Christopher van Eldik**
Erlangen Centre for Astroparticle Physics, Department Physik, Friedrich-Alexander-Universität Erlangen-Nürnberg

Daten werden als entscheidende Ressource des 21. Jahrhunderts betrachtet[1]. In der Physik gibt es eine große Bandbreite von wohlstrukturierten, jahrelangen digitalen Datenströmen aus Großexperimenten (z. B. Astronomie, Teilchenphysik) bis hin zu manueller und hochgradig individuell geprägter Datenaufnahme in Laborexperimenten (z. B. Festkörperphysik, Optik). Wären alle Forschungsdaten der Vergangenheit systematisch erfasst und offen zugänglich, könnten unter Umständen völlig neue Erkenntnisse aus deren Analyse resultieren. Dabei sind nicht nur die Rohdaten selbst, sondern insbesondere auch die Metadaten von größter Bedeutung. Letztere beschreiben Messaufbau, experimentelle Parameter, Einheiten, Kontext, Versionen, etc. und erzeugen damit ein möglichst vollständiges und nachvollziehbares Bild. Je reichhaltiger und systematischer Metadaten erfasst werden, umso besser wird man aus der Kombination verschiedener Datensätze etwas lernen können.

Als Leitgedanke dieses sogenannten *Forschungsdatenmanagements* dienen die FAIR-Prinzipien: Daten sollen auffindbar (*findable*), offen zugänglich (*accessible*), standardisiert und dialogfähig (*interoperable*) sowie wiederverwertbar (*reusable*) gespeichert werden. Man kann sich die weitere Verwendung vielleicht so vorstellen, dass eine WissenschaftlerIn in 10 Jahren einen Computer durch all unsere Daten surfen lässt und dabei Zusammenhänge entdeckt, die den einzelnen Forschungsgruppen verborgen geblieben sind. Bund und Länder haben mit der groß angelegten Initiative für eine Nationale Forschungsdateninfrastuktur (NFDI)[2] erhebliche Fördermittel bereitgestellt, um die wissenschaftlichen Disziplinen diesbezügliche Konzepte ausarbeiten zu lassen.

**Vermitteln unsere Physikcurricula in dieser sich verändernden Welt die passenden Datenverarbeitungskompetenzen?**

In der Physik besteht breiter Konsens, dass die Studierenden fundierte Kenntnisse in Analysis, linearer Algebra, etc. haben müssen, um im Verlauf des Studiums physikalische Konzepte zu verstehen. Man hat sich hier weltweit auf einen gewissen Kanon geeinigt. Unser Eindruck ist, dass ein solcher Konsens bezüglich Datenkompetenzen nicht existiert. Beim Erstkontakt mit Daten, also insbesondere in den Praktika der Experimentalphysik, werden zwar elementare Datenevaluation und systematische Dokumentation eingeübt, diese genügen aber nicht den steigenden Ansprüchen der Forschung und der Berufspraxis, sich zunehmend mit der Analyse großer Datenmengen zu befassen. Es scheitert oft schon an einem Konsens über eine geeignete Einstiegsprogrammiersprache.

Am Department Physik der Friedrich-Alexander-Universität Erlangen-Nürnberg haben wir in den letzten Jahren kleinere Anpassungen im Physikcurriculum vorgenommen, die wir in diesem Artikel vorstellen. Datenkompetenz wurde früh im Bachelorstudium platziert, woraus sich erhebliche Vorteile für den weiteren Studienverlauf ergeben haben. Die Autoren können sich des Eindrucks nicht erwehren, dass die Studierenden in puncto Datenkompetenz schnell auf die Überholspur gehen; wir sehen bereits jetzt in unseren Arbeitsgruppen, dass sie sich als treibende Kräfte hin zu einem modernen Forschungsdatenmanagement erweisen.

---

[1] https://ec.europa.eu/info/sites/default/files/communication-european-strategy-data-19feb2020_en.pdf
[2] https://www.nfdi.de/

**Obligatorisches Datenverarbeitungspraktikum**

Wie führt man Physikstudierende möglichst frühzeitig an computergestützte Datenauswertung heran? Wie begegnet man der Herausforderung, dass manche Studierende bereits aus der Schule oder durch eigene Aneignung Programmiererfahrung vorweisen können, für andere das Thema aber wie ein Buch mit sieben Siegeln erscheint?

Unsere persönliche Erfahrung ist, dass sich das Erlernen einer Programmiersprache am einfachsten durch begleitetes Learning-by-doing anhand physikalischer Fragestellungen gestalten lässt. Die intensive eigene Beschäftigung in Form kleinerer Programmieraufgaben fördert ein tiefes Verständnis und legt die Grundlage für die anstehenden Herausforderungen in der Datenauswertung. Das Bearbeiten von Aufgaben in Teams stärkt den gerade zu Beginn des Studiums so wichtigen Austausch mit KommilitonInnen und fördert früh den Aufbau einer wissenschaftlichen, lösungsorientierten Diskussionskultur. Dabei dürfen vor lauter Programmieren die Inhalte des Fachstudiums nicht in den Hintergrund treten. Nur wenn der Anwendungsbezug klar ersichtlich ist, werden sich ein nachhaltiger Kompetenzgewinn und die Motivation einstellen, das Gelernte im weiteren Studienverlauf zu nutzen.

Wir haben in diesem Sinne – erstmals zum Wintersemester 2018/19 – ein verpflichtendes Computerpraktikum „Datenverarbeitung in der Physik" (DV-Praktikum) in das Curriculum unseres Bachelor-Studiengangs aufgenommen. Im DV-Praktikum werden unter Anleitung von Tutoren Programmieraufgaben im Team bearbeitet und anschließend elektronisch in Form eines fertigen Programms eingereicht und bewertet. Begleitet wird das wöchentliche zweistündige DV-Praktikum von einem interaktiven Online-Kurs, durch den die Studierenden genau die Grundlagenkenntnisse erwerben, die für die Bearbeitung der nächsten Praktikumseinheiten benötigt werden. Dabei beschränken sich die Kursinhalte nicht auf das Erlernen einer Programmiersprache. Das Hauptaugenmerk liegt vielmehr in der Vermittlung grundlegender Datenverarbeitungs- und Datenanalyse-Kompetenzen. So enthält der Kurs beispielsweise einführende Module in Messunsicherheiten und Fehlerfortpflanzung, Interpolation, Fitten von Daten, Statistik-Funktionen und Monte-Carlo-Simulationen, Fourier-Transformationen, numerische Integration, Lösung von Differentialgleichungen sowie grafische Aufbereitung von Ergebnissen. Auch ein einführendes Kapitel zu „Guter wissenschaftlicher Praxis" ist Teil des Kurses. Alle diese Aspekte sind selbst für computeraffine Studierende neu und fordern damit alle Teilnehmenden gleichermaßen.

Als Programmiersprache kommt Python zum Einsatz. Die Grundlagen von Python sind auch für Anfänger schnell zu erlernen, und die Erfahrung zeigt, dass gute Python-Kenntnisse das Erlernen anderer gebräuchlicher Sprachen (z.B. C++, Java) erheblich vereinfachen. Wichtiger noch für unsere Zwecke: Neben der umfangreichen Python-Standardbibliothek stehen mit „numpy"[3], „scipy"[4] und „matplotlib"[5] frei zugängliche und gut kuratierte Basis-Bibliotheken zur numerischen Datenverarbeitung, Statistik und grafischen Aufbereitung zur Verfügung, die in der Physik-Community und weit darüber hinaus zur Anwendung kommen. Im DV-Praktikum werden konsequent Jupyter-Notebooks[6] verwendet; sie ermöglichen, Python-Code innerhalb eines Webbrowser-Fensters zu erstellen und laufen zu lassen. Dabei werden Python-Code und Ergebnisse (Text und Grafiken) gemeinsam in einem Dokument dargestellt und gespeichert und können auf diese Weise sehr einfach mit anderen Interessierten geteilt werden. Jupyter-Notebooks verfügen zudem über dieselben umfangreichen (Text-)Formatierungs-Werkzeuge, die in Wikis zur Anwendung kommen, inklusive Formelsatz, Tabellen und Einbindung von Bildern (siehe Abb. 1).

---

[3] https://numpy.org
[4] https://scipy.org
[5] https://matplotlib.org
[6] https://jupyter.org

Nach einer sehr erfolgreichen Pilotphase haben wir das DV-Praktikum auf besonderen Wunsch der Studierenden bereits im ersten Fachsemester verankert. Um dafür Platz im dicht gedrängten Stundenplan zu schaffen, beginnt das Grundpraktikum deshalb bei uns erst im zweiten Semester. Dies bietet für unsere Studierenden eine perfekte Gelegenheit, die erworbenen Datenverarbeitungs-Kompetenzen gewinnbringend nicht nur bei der Auswertung der Praktikumsversuche einzusetzen, sondern auch im gesamten Studium.

**Spielerischer Zugang zur Physik in Kursvorlesungen**

Mit der Entscheidung für die Programmiersprache Python und der Einbindung ins Erlanger Physikcurriculum kann man in Vorlesungen, Übungen und Praktika auf diese Kompetenzen zurückgreifen. Mit jeder Anwendung wird dieser Kenntnisstand weiter vertieft, das gilt gleichermaßen für Studierende und Lehrende. Das ermöglicht völlig neue Aufgabentypen, bei denen man komplexere, manchmal auch realitätsnähere oder aktuellere physikalische Probleme behandeln kann, als wenn man nur auf analytische Kompetenzen zurückgreifen könnte. Jeder Dozent kann für sich selbst entscheiden, wo analytische Verfahren und wo numerischen Verfahren didaktisch sinnvoll sind.

Erhebliche Vorteile hat die Numerik bei der Visualisierung. Dies ermöglicht eine zweite Herangehensweise an physikalische Probleme: die spielerische. Hierbei lassen sich Parameter schnell variieren und die Ergebnisse in verschiedenen grafischen Auftragungen darstellen – in dieser Qualität und Geschwindigkeit ist dies mit traditionellen Methoden innerhalb einer Übung nicht zu schaffen. Bei manchen wird dabei Neugierde geweckt. Beispiel Fourier-Analyse: Was passiert, wenn man im Signal nur gerade Terme zulässt, oder nur ungerade? Man würde sich wünschen, dass Studierende befähigt sind, solche Fragestellungen sowohl analytisch als auch numerisch zu lösen. Besonders profitieren die Studierenden, die den Vergleich anstellen.

Ein anderes Beispiel betrifft das Misstrauen aufmerksamer Studierender gegenüber Näherungen, die für analytische Lösungswege häufig unvermeidbar sind. Numerisch quantifiziert und visualisiert lässt sich oft viel leichter nachvollziehen, ob und unter welchen Umständen diese Näherungen gut begründbar sind. Wir sind der Überzeugung, dass Lehrveranstaltungen der experimentellen und der theoretischen Physik gleichermaßen von dieser Art spielerischer Datenkompetenz profitieren.

**Elektronisches Laborbuch im Praktikum**

Für ein zukunftsweisendes Forschungsdatenmanagement ist das traditionell handgeschriebene Laborbuch nicht geeignet. Die nachhaltige Verfügbarkeit von Daten und Metadaten lässt sich nur durch sorgfältige elektronische Protokollierung z. B. in Form eines elektronischen Laborbuchs erreichen; so wird von Beginn an eine strukturierte Datenerfassung ermöglicht.

Unsere Universität hat sich im Rahmen eines Pilotprojektes für ein open-source elektronisches Laborbuch (ELN) entschieden (openBIS[7]; siehe Infobox). Dieses wird aktuell in verschiedenen Forschungsgruppen ausprobiert und evaluiert.

---

[7] https://openbis.ch/

Wir haben uns entschlossen, als Vertiefung des DV-Praktikums dieses ELN-System im Physikstudium einzuführen. Geeignet ist dafür das bestehende Elektronikpraktikum mit seinen 11 Versuchstagen im 4. Fachsemester. Dieses bearbeitet ein Programm von einfachen elektrischen Messungen über komplexer werdende analoge und digitale Schaltungen bis hin zu Microcontrollerprogrammierung und Analog/Digital-Wandlung. Bislang gaben wir eine Struktur vor, die Messdaten auf vorgegebenen Pfaden eines Netzwerklaufwerks zu speichern, wohingegen die Experimentvorbereitung und die Metadaten sorgfältig, aber ohne Formvorgaben in einem gebundenen Papier-Laborbuch dokumentiert wurden. Im Frühjahr 2022 vollzogen wir den Übergang zum ELN, der technisch gesehen relativ einfach war, da die Arbeitsplätze sowieso mit einheitlichen Computern ausgestattet waren. Die Lernziele und Inhalte des Elektronikpraktikums blieben unberührt.

Wir haben nur einige wenige Vorgaben implementiert: Beim ersten Login sehen die PraktikumsteilnehmerInnen ihren Laborbuch-*space*, auf den die 2er-Teams gemeinsamen Zugriff haben. Innerhalb dessen sind bereits alle Versuchstage als *projects* und darunter die einzelnen Experimente und Aufgaben als *experiments* und *experimental steps* angelegt. Die kursiv gesetzten Begrifflichkeiten sind von openBIS vorgesehene Standardelemente. Die gesamte Struktur wird von uns per Python-Script angelegt.

Die elektronischen Laborbuchseiten (*experimental steps*) enthalten drei vorgegebene Freitextfelder für (i) die Ziele des Experiments, (ii) die Experimentbeschreibung und (iii) die Dokumentation der Resultate. Messdaten können bei kleineren Messreihen entweder per Hand in Tabellen erfasst werden oder als Rohdatendatei in das ELN geladen werden. Auf diese Weise werden beschreibende Metadaten und Rohdaten von den Studierenden während des Experimentierens an einem Ort zusammengeführt (siehe Abb. 2). Der besondere Vorteil des ELN wird bei Nutzung der Pythonschnittstelle deutlich. Aufbauend auf ihre mittlerweile dreisemestrige Pythonkompetenz können unsere Studierenden hiervon unmittelbar profitieren. Mit einer einzigen Programmzeile lassen sich die Daten direkt aus dem ELN zur weiteren Auswertung und Visualisierung laden (vgl. Abb. 3). Das geht im Praktikum genauso wie zu Hause. Hierzu wurde ein Pythonmodul[8] zur Verfügung gestellt, das Befehle der PyBIS-Bibliothek kombiniert.

Die Studierenden fanden schnell heraus, wie man Fotos von ihrem Smartphone in das ELN einbinden kann. So wurden handgezeichnete Skizzen oder dokumentierende Aufnahmen der Versuchsaufbauten integriert. Das gleiche gilt natürlich auch für Screenshots relevanter Passagen aus externen Dokumenten, z. B. Benutzerhandbüchern von Messgeräten. Die Studierenden haben das elektronische Laborbuch sofort akzeptiert und ausschließlich genutzt, obwohl die Verwendung des ELNs in der Pilotphase freigestellt war.

Natürlich bietet dieses neue Medium auch Gelegenheit, die Arbeitsabläufe sinnvoller zu gestalten. Ein Beispiel: Die Praktikumsvorbereitung erfolgte bislang basierend auf einem Fragenkatalog im Laborbuch. Im Verlauf des Praktikums wurde diese zusammenhängende Darstellung aufgrund der chronologischen Notation in einem Papierbuch oft nicht mehr beachtet. Im ELN lassen sich Vorbereitung und Experimentdurchführung viel besser verzahnen, in dem die Antworten auf die Vorbereitungsfragen und zusätzliche Notizen in die entsprechenden Freitextfelder (Ziele und Experimentbeschreibung) der entsprechenden Experimente eingetragen werden. Vorbereitung, Fragenkatalog und Antworten der Studierenden erscheinen am Ende in einer wohlgeordneten Darstellung, die für die Studierenden sehr viel besser lesbar ist. Es sei auch erwähnt, dass die Betreuer sich durch ein Pythonskript eine exzerpierte Version der Vorbereitung ansehen können, um sich einen Überblick über den Vorbereitungsstand der Praktikumsgruppen zu verschaffen, ohne sich immer wieder durch die einzelnen Experimente im ELN klicken zu müssen.

---

[8] https://github.com/FAU-PHYSIK-EP/load_openBIS_data_into_python

Gegenüber traditionellen Laborbüchern erlaubt ein ELN, Nachbesserungen, Plagiate oder Fälschungen leichter zu verhindern. Die Studierenden wissen natürlich bereits von ihrer Internetnutzung, dass jede Handlung in einem elektronischen System prinzipiell genau nachvollziehbar sein könnte. Somit sollte gar nicht erst der Gedanke aufkommen, Daten zu "frisieren".

Die Betreuer haben wiederum über den Rohdatenzugriff eine Möglichkeit, die Versuchsaufbauten im Verlauf der Zeit bei etwaigen Ergebnisabweichungen zu überprüfen und so technischen Defekten auf die Schliche zu kommen.

Nach Abschluss einer Lehrveranstaltung sollten Studierende eine vollständige Dokumentation für sich behalten können. Neben der Weiternutzung des ELN im Fortgang des Studiums gibt es auch die Möglichkeit, die ELN-Inhalte zu exportieren.

**Wirkung auf die Forschungsebene**

Nachdem die Studierenden den Umgang mit dem ELN im 4. Semester erlernt haben, bringen sie ihre Kompetenzen während der Forschungsphase zur Bachelor- oder Masterarbeit unmittelbar in die Arbeitsgruppen ein. Dort muss die Erfassung der Forschungsdaten allerdings wesentlich strukturierter und unmittelbar an die Arbeitsabläufe in der jeweiligen Forschungsgruppe angepasst erfolgen. Diese Aufgabe steht der großen Mehrheit noch bevor und ist Gegenstand der NFDI-Initiative. Auch wir sind gerade dabei, unsere Arbeitsabläufe in der Forschung systematisch zu erfassen und abzubilden. Dabei helfen Konzepte aus der Informatik, z. B. Entity-Relationship-Diagramme, wie sie für Datenbanken erstellt werden. Hier hilft der ELN-Einsatz im Praktikum: Wir werden Konzepte mit stärker strukturierten Laborbüchern zunächst im Praktikum erproben und evaluieren. Aufgrund der abgeschlossenen Praktikumseinheiten sind dabei kurze Lern- und Innovationszyklen möglich, deren Erfahrungen wir dann in die längerfristig angelegte und komplexere Dokumentation der Forschung in unserer Arbeitsgruppe einbringen können.

Wir sehen es als ein Erfolgsrezept, frühzeitig im Physikcurriculum zeitgemäße Datenkompetenz zu vermitteln. Die durchgängige Verwendung einer einheitlichen und in den Naturwissenschaften mächtigen Programmiersprache wie Python verschafft neue Möglichkeiten und verändert die Lehre und die Forschung nachhaltig. Wir sehen einen unmittelbaren Einfluss auf den Übungsbetrieb und den Praktikumsbetrieb im Bachelor- und Masterstudium sowie zusätzliche Kompetenzen der Studierenden, die sie in der Forschungsphase zur Bachelor- und Masterarbeit in die Arbeitsgruppen einbringen können. Darüber hinaus sehen wir einen sehr positiven Einfluss auf die sich verändernde Forschungslandschaft, die viel stärker als früher von einem aktiven Forschungsdatenmanagement geprägt sein wird.

Heiko B. Weber und Michael Krieger forschen am Erlanger Department Physik an festkörperphysikalischen Fragestellungen, leiten das Elektronikpraktikum und arbeiten aktiv im NFDI-Konsortium FAIRmat mit.

Christopher van Eldik forscht am selben Department an astroteilchenphysikalischen Fragestellungen. Als Studiendekan hat er die Umstellung des Curriculums mitgeprägt und leitet die Lehrveranstaltung "Datenverarbeitung in der Physik".

Abbildungen:

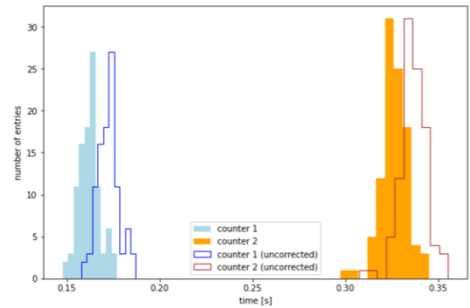

Abb. 1: Beispielanalyse zur Bestimmung der Fallbeschleunigung aus dem Datenverarbeitungspraktikum im 1. Fachsemester. In solchen interaktiven Jupyter-Notebooks werden Python-Code, Dokumentation und Abbildungen auf einer Ebene dargestellt.

Abb. 2: Ansicht des ELN während des ersten Versuchstages im Elektronikpraktikum: in der Baumansicht haben wir den Praktikumsablauf vorgezeichnet. Die Studierenden haben während der Vorbereitung auf das Praktikum bereits Ziele und Durchführung der einzelnen Praktikumsexperimente (Teilaufgaben) inklusive Skizzen in die entsprechenden Eingabefelder notiert. Ergebnisse und Beobachtungen werden während des Praktikums im gleichen Formular erfasst. Hierfür stehen ein Freitextfeld, eine Tabelle für numerische Messdaten sowie der Dateiupload für Messdatendateien zur Verfügung.

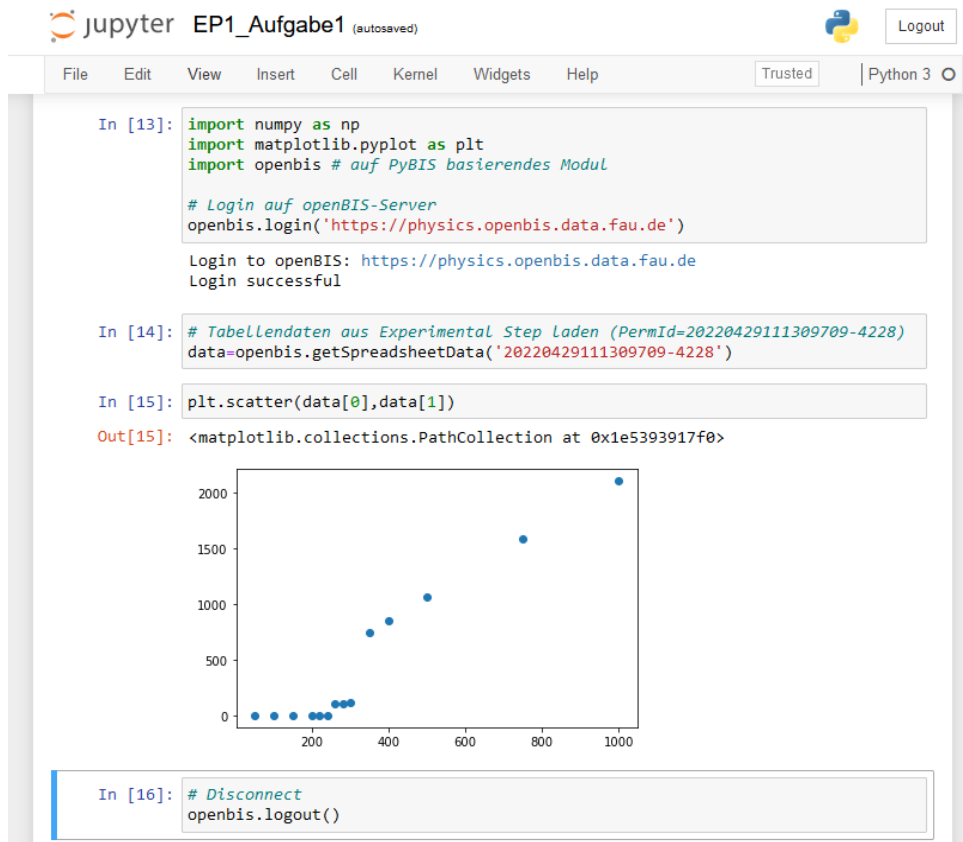

Abb. 3: Grafische Visualisierung von Messdaten aus dem Elektronikpraktikum (vgl. Abb. 2); mit wenigen Programmzeilen laden die Studierenden ihre im Praktikum erfassten Messdaten direkt aus dem ELN in ein Jupyter-Notebook, wo sie analysiert werden.

**Infobox**

Electronic Lab Notebooks (ELN)

Unter den verfügbaren open-source Softwarelösungen hat sich unsere Universität in der Pilotphase für openBIS entschieden. Ein anderes in der Physik gängiges open-source System wäre elabFTW. openBIS bringt viele Funktionalitäten mit sich und erlaubt, Datenstrukturen ähnlich einer Datenbank zu definieren und zu verknüpfen. Dabei gilt es für den Nutzer, die Balance zwischen Effizienz (einfache Formulare; Freitexteingabe) und Perfektion (experimentspezifische Formulare und resultierende strukturierte Metadatenablage), Schnittstellen zu Forschungspartnern, etc zu finden. Dies ist zu einem großen Teil eine aktuelle und andauernde Forschungsfrage in den verschiedenen NFDI-Konsortien (z. B. FAIRmat).

Eine wichtige Funktionalität eines ELN-Systems ist die Exportschnittstelle. openBIS bietet hier sowohl die Möglichkeit, gut lesbare, formatierte Dokumente zu erstellen als auch strukturierte Daten (JSON-Format) zu exportieren – diese Interoperabilität ist eine Grundvoraussetzung für die Umsetzung der FAIR-Prinzipien.